# Pragmatic Earth-Fixed Beam Management for 3GPP NTN Common Signaling in LEO Satellites

**Xavier Artiga[1], Màrius Caus[1], Ana Pérez-Neira[1,2]**
[1]Centre Tecnològic de Telecomunicacions de Catalunya (CTTC/CERCA), Castelldefels (Barcelona), Spain
[2]Universitat Politècnica de Catalunya (UPC), Barcelona, Spain

Corresponding author: Xavier Artiga (e-mail: xavier.artiga@cttc.es).

This work was supported by the project 5G-STARDUST, which has received funding from the Smart Networks and Services Joint Undertaking (SNS JU) under the European Union's Horizon Europe research and innovation programme under Grant Agreement No 101096573 and; by the project SOFIA PID2023-147305OB-C32 funded by MICIU/AEI/10.13039/501100011033 and by FEDER/UE.

**ABSTRACT** This work proposes a pragmatic method for the design of beam footprint layouts and beam hopping illumination patterns to efficiently broadcast 3GPP NTN common signaling to large coverage areas using EIRP-limited LEO satellites. This method minimizes the time resources required to sweep over the whole coverage while ensuring that the signal-to-interference-plus-noise ratio received by users is above a given threshold. It discusses the design of: (i) an Earth-fixed grid of beam layouts; (ii) beamforming vectors and beam power allocation; (iii) beam hopping patterns and (iv) space, time and frequency resource allocation of 3GPP common signaling. Two main beam layout solutions are proposed to significantly reduce the number of beams required to illuminate the coverage area: one based on phased array beams with low beam crossover levels and the other on widened beams. A numerical evaluation using practical system parameters showed that both solutions perform similarly, but that the best result is obtained with phased arrays beams with optimized beam cross over levels. Indeed, for the system evaluated, they allowed reducing the total number of beams from 1723 to 451, which combined with a proper beam hopping pattern and scheduling scheme allowed obtaining a coverage ratio of 100% and a common signaling efficiency (i.e. number of slots carrying common signaling over total number of slots) up to 80.6% for the most stringent common signaling periodicity of 20 ms considered by 3GPP.

## I. INTRODUCTION

Low Earth Orbit (LEO) mega constellations represent a paradigm shift for satellite communication systems. Satellite beam management is one of the system aspects requiring a new approach. High throughput satellites deployed in Geostationary Earth Orbits (GEO) usually relied on multispot beam layouts fixed from both the satellite and the Earth point of views, due to the relative quasi static position of the satellite. Large GEO platforms allow the simultaneous illumination of all beams within the satellite field of view (FoV), so frequency reuse schemes based on multiple colors are used to contain inter-beam interferences. In contrast, smaller LEO platforms become EIRP (Equivalent Isotropic Radiated Power) limited, so they cannot simultaneously illuminate all beams to cover a large region, as sketched in Fig. 1.

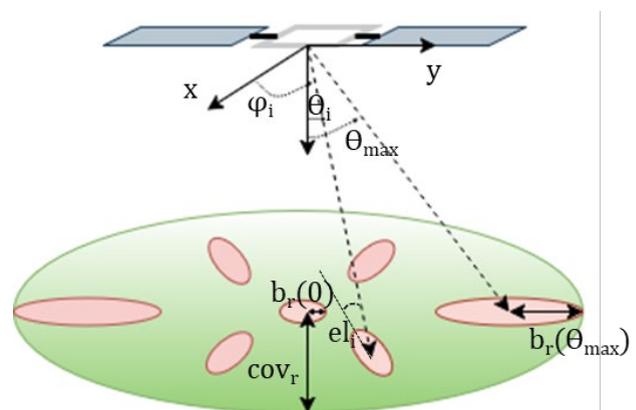

**FIGURE 1.** Sketch of an EIRP-limited LEO system. Only a few beams can be simultaneously illuminated within the coverage area.

Commercial megaconstellations address this challenge in different ways [1]. Eutelsat-Oneweb relies on a train of highly elliptical Earth-moving beams (i.e. fixed from the satellite point of view) resulting in a reduced FoV. Starlink, Telesat and Amazon Kuiper [2] use steerable antennas capable of synthesizing Earth-fixed beams, but they can only simultaneously illuminate a very limited number of them, within a much larger FoV. In this context, beam hopping solutions to cover the whole FoV have not been reported.

### A. RELATED WORKS

Recently, the interest on the 3$^{rd}$ Generation Partnership Project (3GPP) New Radio (NR) standard, which supports non terrestrial networks (NTN) since Release 17, has been reinforced since it is postulated as the communication technology for next European constellation IRIS$^2$ [3]. So far, for NTN, the 3GPP standard adopted the beam management framework designed for terrestrial systems allowing the alignment of satellite and user terminal beams. Other issues like beam to cell mapping, beam layout design, beam hopping or frequency reuse schemes are left out of the standard [4]-[5]. Only an Earth-moving beam layout and three different frequency reuse schemes were recommended to perform preliminary system level analysis [6]. The most stringent requirement of 3GPP standard for the design of a beam layout and a beam hopping pattern is the periodicity in which certain control signaling common to all users must be transmitted. In Release 17 user terminals must assume that Synchronization Signal Blocks (SSB) were transmitted with a 20 ms periodicity [7], which implied that the beam hopping pattern must sweep over the whole coverage in less than 20 ms. This triggered a discussion on the flexibilization and increase of such periodicities [8] or the reduction on the total number of beams covering the FoV by reducing the beam crossover levels [9]. Finally, in Release 19 an SSB extended periodicity of 160 ms is additionally supported for NTN [10].

The 3GPP standard supports both Earth-fixed and moving beams [11]-[12]. However, from a system level point of view, Earth-fixed solution is preferred since it does not require frequent handovers due to satellite movement. The price to pay is the increased complexity in the beam management. First, beamforming weights need to be frequently updated to keep the beam pointing to a given Earth position. Second, the beam footprint over a specific position changes depending on the distance of this position to the sub satellite point, mainly due to the Earth curvature [2], [11] and [13] . The consequence is that to realize a beam layout without coverage *holes*, the fixed grid over the Earth must be designed according to the beam footprint radius at Nadir, leading to a large total number of beams, which aggravates the beam hopping issues discussed above, and a large beam overlapping at coverage edges [13]. Alternatively, beam widening techniques can be used to reduce both the number of beams and the overlapping at the cost of a reduced EIRP per beam [2], [11] and [13].

Beam widening has been well studied in the literature (e.g. [14]-[17]). There are three main solutions: aperture truncation (i.e. switching off some elements making the resulting antenna aperture smaller), amplitude tapering and phase tapering. In aperture truncation and amplitude tapering not all amplifiers are used at their maximum power resulting in a reduction on the total transmit power. In contrast, phase tapering keeps transmit power at the price of increased gain ripples. Trading off beamwidth, gain ripples and sidelobe levels requires complex non-convex optimization not always providing global optimal solutions [16].

Besides EIRP limitation and Earth-fixed footprint challenges, the beam management solution must also deal with the different requirements of common signaling and data transmission. Efficient common signaling needs a beam sweep over the whole coverage using the minimum amount of time and frequency resources, which may favor the use of reduced number of wide beams. In contrast, system capacity is increased if high gain narrow beams are directed to a single user or a cluster of users, leading to user-centric beamforming schemes for data transmission [4]. This dual set of requirements was addressed by resorting to the combination of a wide beam for signaling and spot beams for data transmission [18]. However, using a single wide beam to cover the entire FoV would not provide sufficient EIRP to close the link in practice. As a result, synthesizing multiple wide beams together with an appropriate beam-hopping pattern remains an open research challenge. Alternative approaches have proposed more complex beam-hopping schemes that illuminate signaling and data beams simultaneously [19], but these do not leverage the distinct beam footprints required for the two functionalities. In addition, they address only SSB transmissions on the signalling side.

### B. MAIN CONTRIBUTIONS

This study proposes a 3GPP NTN beam management framework for Earth-fixed beam systems. It assumes two sequential stages: (i) the transmission of broadcast beams for common signaling; and (ii) the transmission of data beams based on user centric solutions for user-specific data and control information. Since data transmission according to demand distribution has been object of multiple studies, including beam hopping mechanisms (e.g. [20] and references therein), here the focus is on the design of broadcast beams. The main objective is then to design a multibeam layout, a beam hopping illumination pattern and a scheduling strategy for the different 3GPP common control signals to minimize the number of time resources devoted to common signaling, thus maximizing system capacity for data transmission. The focus is on the downlink since satellite EIRP limitation does not apply to uplink. To the best of the authors knowledge, this is the first work to report such an end-to-end approach (spanning from beam design to signaling scheduling) for 3GPP NTN systems.

### C. ORGANIZATION

The rest of the paper is organized as follows. Section II analyzes 3GPP NTN common signaling and derives the corresponding scheduling requirements. Section III describes the system model. Section IV details the proposed methods for the design of the beam layout. Section V discusses a regular time multiplexing scheme for beam illumination. Section VI performs a numerical evaluation of the proposed solutions using practical system parameters. Section VII addresses the common signaling spatial, time and frequency allocation completing the beam hopping pattern design. Finally, Section VIII concludes the work.

## II. 3GPP NTN COMMON CONTROL SIGNALING

In the downlink, common signaling refers to a set of control signals broadcasted by the satellite to all users enabling key operations such as initial cell attachment and reception of paging messages indicating incoming traffic. More specifically, common signaling includes [7]:

### 1) SYNCHRONIZATION SIGNAL/PBCH BLOCK (SSB)

The SSB is used by the User Equipment (UE) to acquire downlink synchronization and to get the Master Information Block (MIB). It is confined within a 5 ms window and the default periodicity is 20 ms in Release 17. Recent 3GPP discussions on coverage enhancement [8]-[9], [21], resulted in the inclusion of an alternative value of 160 ms for NTN in Release 19. The SSB spans 240 subcarriers (20 resource blocks) and 4 OFDM symbols. The SSB carries two synchronization signals, i.e., primary synchronization signal (PSS) and secondary synchronization signal (SSS), as well as the physical broadcast channel (PBCH), which contains the MIB, that in turn provides information on how to detect the System Information Block 1 (SIB1).

### 2) PHYSICAL DOWNLINK CONTROL CHANNEL (PDCCH)

The PDCCH provides the time and frequency allocation for common signaling transmissions using downlink control information (DCI) messages. Each DCI message is scrambled with a specific radio network temporary identifier (RNTI) that depends on the usage. For instance, system information (SI), random access (RA) and paging (P) messages are respectively assigned to SI-RNTI, RA-PNTI and P-RNTI codes. The time and frequency allocation of PDCCH transmissions are given by a set of configured search spaces. UEs monitor Search Space Type 0 and 0A using SI-RNTI for allocation of SIB1 and other SIBs like SIB19, respectively; Type 1 using RA-RNTI for random access (RA) MSG2 and MSG4; and Type 2 with P-RNTI for paging messages. The configuration of Search Space Type 0 is included in MIB, whereas Type 0A, 1 and 2 are included in SIB1. Search Spaces are mapped to a specific control resource set (CORESET), which defines the set of contiguous resource blocks and OFDM symbols that may carry the PDCCH. The number of resource elements (REs) of a CORESET that are required to carry a PDCCH DCI message is called aggregation level, which is expressed in terms of control channel elements (CCEs). SIB1 allocation is carried within CORESET0, which is defined by the 3GPP standard and depends on the SSB-CORESET0 multiplexing pattern.

### 3) SYSTEM INFORMATION BLOCK 1 (SIB1)

SIB1 is the first broadcast message that transmits cell-specific information. It is essential for proper network operation. SIB1 periodicity is 160 ms, though repetition periodicity within the 160 ms window is determined by the network implementation. SIB1 provides details about the search space as well as periodicity and windowing for other SIBs, e.g., SIB19. The size of SIB1 depends on the optional information elements included. In this study, we consider 1280 bits, which was agreed by the 3GPP for system level simulations on downlink coverage [21].

### 4) SYSTEM INFORMATION BLOCK 19 (SIB19)

SIB19 carries essential information for NTN operation such as satellite ephemeris. It is transmitted in system information (SI) messages. The periodicities for SIB19 are chosen from {80, 160, 320, 640, 1280, 2560} ms. SIBs having the same periodicity can be mapped to the same SI message if the maximum SI messages size allows. The maximum size of an SI message is 2976 bits. Each SI message is transmitted within a time window, which is referred to as SI-window. The SI-windows of different SI messages do not overlap. To enhance the downlink coverage, each SI message can be repeated within the SI-window. Similarly to the SIB1 definition, we assume the size of SIB19 derived in [21], which is equal to 616 bits.

### 5) RANDOM ACCESS RESPONSE-MESSAGE 2 (MSG2)

The RA procedure consists of four sequential messages, two of them in the downlink direction. MSG2 is the RA response transmitted by the gNB after the successful detection of a RA preamble in a given RA occasion. It carries critical information such as the time advance command for timing adjustment and an initial uplink grant for the UE. It also assigns a temporary identifier (TC-RNTI) to the UE. A single MSG2 is transmitted per RA occasion, so it may carry information for multiple UEs that initiated the random access procedure on the same occasion. The size of MSG2 depends on how many UE are addressed. The resources allocated to MSG2 are estimated assuming that 70 bits per UE are required.

### 6) PAGING

The paging messages are sent from the gNB to UEs in IDLE state to initiate calls. In IDLE state, the gNB does not know the exact position in which the user is located so paging messages are transmitted in a tracking area basis, to ensure they reach the target user. A single paging message can address up to 32 UE, with either the 5G-S-TMSI or the I-RNTI identifier, requiring up to 1536 bits or 1280 bits, respectively.

### 7) RANDOM ACCESS MESSAGE4 (MSG4)

RA MSG4 conveys the contention resolution information that allows a user to confirm that it has been correctly identified by the gNB. The identified user adopts the temporary identifier TC-RNTI as user identifier C-RNTI. Besides, MSG4 may also convey the RRC setup information element, which provides the UE the configuration required to carry out the network registration procedure. Strictly speaking, MSG4 is not a common signaling transmission. However, it is included here since before receiving MSG4, a UE has no dedicated configuration to receive dedicated data transmissions. According to [21], we assume a size of the MSG4 of 1040 bits.

Table I summarizes the required SNR for the proper detection of the different common signals with a Block Error Rate (BLER) of 1% derived in [8]. It is taken as reference to design the beam layout and beam hopping pattern.

TABLE I
SNR REQUIRED FOR A BLER OF 1%

| Signal | Required SNR (dB) |
|---|---|
| SSB | -6.3 |
| PDCCH | -6 |
| PDSCH msg2 | -10.9 |
| PDSCH msg4 | -5.2 |
| PDSCH SIB1 | -5.8 |
| PDSCH SIB19 | -6.9 |
| PDSCH for 1Mbps data rate | -4.1 |

## III. SYSTEM MODEL

Let us consider the downlink transmission of common signaling from a LEO satellite to a coverage area with radius $cov_r$, as sketched in Fig. 1. The coverage area is defined as the region in which the users see the satellite with an elevation angle $el_i \leq el_{min}$. The satellite is equipped with a Direct Radiating Array (DRA) with $N_a$ elements capable of producing a beam footprint (corresponding to the half-power beamwidth) at Nadir with radius $b_r(0)$, with $b_r(0) \ll cov_r$. A quasi Earth-fixed beams scheme is adopted, so a regular grid of beam footprint centers is defined on Earth. As the satellite moves over the Earth, the set of $K(t)$ beam centers included within the coverage area varies [12]. The satellite is assumed to be EIRP-limited, so it cannot simultaneously illuminate the $K(t)$ beams. Therefore, a beam hopping scheme sweeping around the whole coverage is assumed. The transmitted beam hoped signal can be expressed as

$$S(t) = \sum_{ih=0}^{N_h-1} a_{ih}(t) \mathbf{W}_{ih}(t) \mathbf{x}_{ih}(t) \quad (1)$$

where $ih$ is the hop index; $N_h$ the total number of hops; $a_{ih}(t)$ is the activation function

$$a_{ih}(t) = \begin{cases} 1 & t \in T_{ih} \\ 0 & t \notin T_{ih} \end{cases} \quad (2)$$

where $T_{ih}$ designates the time interval of hop $ih$; $\mathbf{W}_{ih}(t)$ is the $N_a \times N_b(ih)$ beamforming matrix in hop $ih$, where $N_b(ih) < K(t)$ is the number of active beams in hop $ih$, which can be gathered in the set $\mathbb{S}_a(ih)$; and $\mathbf{x}_{ih}(t)$ is the $N_b(ih) \times 1$ vector of OFDM transmitted signals, which components are assumed to be statistically independent and normalized so that $E\{|\mathbf{x}_{ih}|^2\} = 1$. The dependency of $\mathbf{W}_{ih}(t)$ on time reflects the need of updating the beamforming weights to keep pointing to the target footprint centers despite the satellite movement. However, the assumption is that the required refreshing period is larger than the hop duration so the beam footprint variations within each hop are negligible. Note that in Section VII hop durations are in the order of 1 slot, corresponding to 125 μs, which translate to pointing errors below 0.001° if beamformers are updated every hop.

The signal received by the $i$-th user terminal in hop $ih$, at instant of time $t$, which is omitted for simplicity based on the previous assumption, is expressed as

$$y_i = \mathbf{h}_i \mathbf{w}_k x_k + \sum_{\substack{j \in \mathbb{S}_a(ih) \\ j \neq k}} \mathbf{h}_i \mathbf{w}_j x_j + z_i \quad (3)$$

where $\mathbf{h}_i$ is the $1 \times N_a$ wireless channel vector associated to user $i$ and $\mathbf{w}_k$ is the $N_a \times 1$ beamforming vector pointing at the $k$-th beam center. In this example, user $i$ is located within the area illuminated by the $k$-th beamforming vector, which is assumed to convey the intended signal $x_k$. Index $j$ in (3) corresponds to all active beams in the current hop and $x_j$ is the signal transmitted by the $j$-th beam. Finally, $z_i$ are the noise samples, modeled as independent and identically distributed random Gaussian variables, i.e., $z \sim \mathcal{CN}(0, \sigma_n^2)$. The component of the channel vector corresponding to the $i$-th user and the $n$-th DRA element is modeled as

$$h_{i,n} = \sqrt{\frac{G_T(\theta_i) G_R(el_i)}{L_{pl}(r_i) L_{at}(el_i)}} e^{\frac{j2\pi}{\lambda} r_i} e^{-\frac{j2\pi}{\lambda}(x_\mathbf{n} u_i + y_\mathbf{n} v_i)} \quad (4)$$

for $1 \leq n \leq N_a$. Let $G_T$ denote the radiation pattern of the elements in the satellite DRA, which is modeled as $G_T(\theta_i) = cos(\theta_i)^2$, where $\theta_i$ the tilt angle from Nadir in which the satellite antenna sees user $i$; $G_R$ is the gain of the user terminal antenna; $L_{pl}$ is the pathloss; $L_{at}$ is the atmospheric loss; $r_i$ is the slant-range; $(x_n, y_n)$ are the coordinates of DRA element $n$; and $u_i = sin(\theta_i)cos(\varphi_i)$ and $v_i = sin(\theta_i)sin(\varphi_i)$ are the angular coordinates of user $i$. Both $L_{at}$ and $G_R$ depend on the elevation angle over the horizon $el_i$ in which the UE sees the satellite (see Fig. 1), to account for the effective path length through the atmosphere and scan loss, respectively. Although the methodology proposed in following sections can be applied in different frequency bands, numerical evaluations focus on Ka-band, i.e. frequency range (FR2). Therefore, multipath effects are considered negligible due to the directivity of the UE antenna. Besides, shadowing effects are not modeled but considered as a link budget margin over the required SNR summarized in Table I.

The Signal-to-Interference-plus-Noise Ratio (SINR) received by the *i*-th UE, assuming that its intended signal is conveyed in the *k*-th beam, is then

$$SINR_i = \frac{|\mathbf{h}_i \mathbf{w}_k|^2}{\sum_{j \neq i}|\mathbf{h}_i \mathbf{w}_j|^2 + \sigma_n^2} \quad (5)$$

Sections IV-VI design a generic beam hopping scheme minimizing the required number of hops to sweep over the whole coverage while ensuring a received SINR above a target threshold $SINR_{th}$. This includes the identification of the $K$ beam footprints required to illuminate the coverage area and the corresponding beamforming matrices $\mathbf{W}$, as well as the derivation of the active beams in each hop $\mathbb{S}_a(ih)$ and the total number of hops $N_h$ required. Let us remark that minimizing the number of hops devoted to common signaling transmission maximizes the time resources that could be devoted to data transmission through user centric beamforming. Afterwards, Section VII addresses the specific scheduling of common signaling leveraging on the designed beam hopping patterns. It defines the hop duration and the number of full beam hopping sweeps across the coverage required to broadcast all common signaling. Moreover, it discussed the impact of the scheduling scheme on beam to NR cell mapping solutions.

## IV. BEAM FOOTPRINT LAYOUT

To design a regular grid of beam footprint centers fixed on Earth we resort to a hexagonal grid with a distance between grid points fulfilling $\sqrt{3}b_r(0) \leq d_g \leq \sqrt{3}b_r(\theta_{max})$, where $b_r$ is the beam footprint radius corresponding to the -3 dB beamwidth, which depends on the tilt angle from Nadir; and $\theta_{max}$ is the maximum tilt angle at coverage edge (see Fig. 1). Unless otherwise stated, all beam footprint radius correspond to the half-power beamwidth. Specifically, four beam footprint designs are considered, as exemplified in Fig. 2:

### 1) DESIGN A: UNIFORM BEAMS

From a system level point of view, it would be desired to design not only a uniform grid of footprint centers but a uniform grid of beam footprints, despite the position of the satellite and the Earth curvature. This requires setting $d_g = \sqrt{3}b_r(\theta_{max})$ and widening the beams so that they all have a similar footprint. Indeed, beams at the edge will need to be widened on the azimuth dimension, while the rest will need widening in both dimensions. The main drawback of this solution is that beam widening implies a reduction of beam gain, being the beam at Nadir the most affected. On the positive side, this solution provides the minimum number of beam footprints to cover the whole coverage.

### 2) DESIGN B: MAXIMUM GAIN BEAMS WITH -3 dB BEAM CROSSOVER

The opposite solution to design A is a beam design targeting maximum beam gain. In this case we set $d_g = \sqrt{3}b_r(0)$, so phased array beams providing maximum gain and minimum beam footprints present crossover levels of -3dB close to Nadir region, ensuring that any user is served within the half power beamwidth of a maximum gain beam. The drawback here is the resulting large number of beams filling the coverage area and the large beam overlapping at coverage edge. This option is assumed as the benchmark not requiring any beam layout optimization.

### 3) DESIGN C: NON-UNIFORM WIDENED BEAMS

A compromise solution between designs A and B is to set the distance between grid points anywhere between the two limits. We consider setting $d_g = \sqrt{3}b_r(\alpha\theta_{max})$, with $\alpha \in (0,1)$, i.e. using as a reference the beam footprint radius at $\alpha$ times the tilt angle at the edge. It requires widening beams presenting tilt angles $\theta_k < \alpha\theta_{max}$ whereas beam overlapping will appear for $\theta_k > \alpha\theta_{max}$.

### 4) DESIGN D: MAXIMUM GAIN BEAMS WITH REDUCED BEAM CROSSOVERS

A simplified compromise solution is the use of the same phased array beams as in Design B but with $\sqrt{3}b_r(0) < d_g < \sqrt{3}b_r(\theta_{max})$ [9]. In this case, the total number of beams and the beam overlapping at coverage edge is reduced at the expense of reduced crossover levels for beam footprints close to Nadir.

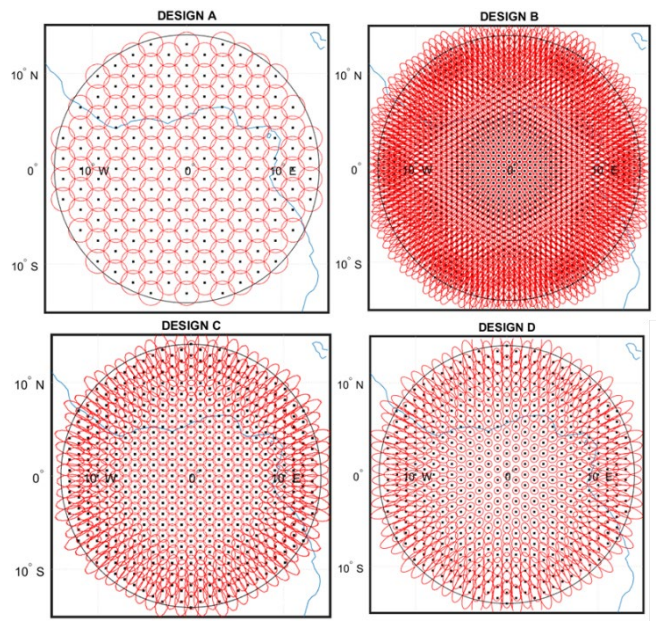

**FIGURE 2.** Beam footprint contours (at half power beamwidth) for the four proposed designs with $b_r(0) = 41.3\ km$, $b_r(\theta_{max}) = 138.6\ km$ and $cov_r = 778,7\ km$. For Designs C and D *dg* is set to $d_g = \sqrt{3}b_r(0.75\theta_{max})$. Black dots correspond to the designed Earth-fixed grid of beam footprint centers.

Let us recall that Fig. 2 shows the beam footprint contours at a given instant. As the satellite moves, grid points at the center of the coverage in Fig. 2 move to the edge and their corresponding beam footprints get deformed for all beam layout designs except Design A, which synthesizes almost uniform beam footprints.

Designs B and D rely on phased array beams. The beamforming vector for the k-th beam is then

$$w_k = \sqrt{\frac{P_k}{N_a}} e^{\frac{j2\pi}{\lambda}(\mathbf{x_n} u_k + \mathbf{y_n} v_k)} \quad (6)$$

where $P_k$ is the power allocated to the *k*-th beam. The procedures for computing the beamforming vectors in the case of widened beams (i.e. Designs A and C) and for calculating $P_k$ are described in next subsections.

### A. BEAM WIDENING

As a beam widening technique we resort to a closed-form phase tapering solution. Phase tapering is selected to preserve the radiated power, since all amplifiers can transmit at their maximum available power. The closed-form expression allows us to get rid of complex non-convex optimizations trading off gain ripple and sidelobe levels that often present multiple local optima [16]. In practice, this limits the possibility of having a systematic approach able to achieve the ripple and sidelobe level requirements for different beam synthesis problems. Moreover, the computational complexity of running non-convex optimizations for each beam as the satellite moves over the Earth-fixed grid is not negligible. Therefore, we follow the approach in [17], leveraging on the fact that the Fourier transform of a complex linearly frequency modulated pulse can be approximated by a rectangular window. Specifically, the *k*-th beamforming vector writes as

$$\begin{aligned}w_k &= \sqrt{\frac{P_k}{N_a}} e^{\frac{j2\pi}{\lambda}(\mathbf{x_n} u_k + \mathbf{y_n} v_k)} \\ &\circ e^{\left(i\frac{2\pi}{\lambda}\left(\frac{x_r^2 \cos(w_x)\sin(w_y) + y_r^2 \sin(w_x)}{D}\right)\right)}\end{aligned} \quad (7)$$

where ∘ represents the Hadamard product. The first term of $w_k$ is the traditional phased array beamformer whereas the second term corresponds to the phase tapering, with $w_x$ and $w_y$ being the target angular widening in the DRA *x* and *y* dimensions. Note however that we not only need to apply beam widening in the principal axes of the antenna but in the angles determined by the position of the footprint centers ($\theta_k, \varphi_k$). To achieve this, and in contrast to [17], we apply the phase tapering to a rotated version of the antenna positions, so that the principal axes of the rotated positions align with the required widening directions. The rotated positions are calculated as

$$\begin{aligned}x_r &= \mathbf{x_n}\cos(\varphi_k) - \mathbf{y_n}\sin(\varphi_k) \\ y_r &= \mathbf{x_n}\sin(\varphi_k) + \mathbf{y_n}\cos(\varphi_k)\end{aligned} \quad (8)$$

The step-by-step process for calculating the beamformer for all widened beams in designs A and C is as follows:
1) Calculate the contour at -3 dB of the reference beam footprint at $\theta_{max}$ or $\alpha\theta_{max}$.
2) Calculate the radius $b_r$ of the beam footprint corresponding to the $\varphi = 0°$ axis.
3) Calculate the distance between centers of the hexagonal grid of beams on Earth as $d_g = \sqrt{3}b_r$.
4) Generate the hexagonal grid and create circles of radius $b_r$ centered in each grid point.
5) Transform the circles from geodetic coordinates to u-v plane.
6) Calculate $w_x$ and $w_y$ as the lengths of the major and minor axes of the resulting ellipses.
7) Apply (5) to calculate the beamforming weights.

### B. BEAM POWER ALLOCATION

Power allocation to simultaneous transmitted beams in each hop *ih* can be used to equalize pathloss and/or directivity differences due to beam widening. Here we compare two different schemes: (i) equal power allocation without equalization, i.e. $P_k = P_a N_a / N_b(ih)$, where $P_a$ is the available power at each DRA element; (ii) and power allocation for SNR equalization. The latter is intended to compensate for SNR differences coming from both the channel and the beam widening process. In this case, the power per beam is computed as

$$P_k = P_a N_a \frac{c_k}{\sum_{j \in \mathbb{S}_a(ih)} c_j} \quad (9)$$

with

$$c_k = \frac{trace(|\mathbf{HW}|^2)}{|\mathbf{hc}_k \mathbf{w}_k|^2} \quad (10)$$

being $\mathbf{hc}_k$, the channel vector of a reference user located at the *k*-th beam footprint center; and $\mathbf{H}$ and $\mathbf{W}$ the corresponding channel and beamforming matrices for the active beams within a hop:

$$\begin{aligned}\mathbf{H} &= \left[\mathbf{hc_1}^T, \mathbf{hc_2}^T, \dots, \mathbf{hc}_{N_b(ih)}^T\right]^T \\ \mathbf{W} &= \left[\mathbf{w_1}, \mathbf{w_2}, \dots, \mathbf{w}_{N_b(ih)}\right]\end{aligned} \quad (11)$$

The term $c_k$ is inversely proportional to the power received by a terminal located in the *k*-th beam center; hence more power is allocated to beams experiencing worse channel conditions or large beam widenings.

### V. BEAM MULTIPLEXING

The aim of this section is to identify the set of active beams in each hop, so $\mathbb{S}_a(ih)$. Three conditions are imposed:
1) Active beam footprints must not overlap to reduce inter-beam interference.
2) One beam is only active in one hop within the coverage sweep, so all beams are fairly illuminated.
3) The multiplexing solution must not depend on the satellite position to avoid complex calculations as the satellite move over the Earth grid.

The proposed solution consists of assigning a hop index *ih* to each point in the Earth-fixed grid. All points within the coverage area sharing the same hop index are served simultaneously in the same hop, whereas points with different hop index are time multiplexed in different hops. The resulting number of hops required to sweep over the whole coverage $N_h$ corresponds to the total number of hop indexes used. To fulfill conditions 2) and 3), a regular hop index assignment guaranteeing maximum distance in both latitude and longitude dimensions between beam footprints within the same hop is proposed. It maps consecutives hop indexes from $ih = 0$ to $ih = N_h - 1$ to rectangular blocks with a number of columns ($nCols$) and a number of rows ($nRows$) equal to

$$nCols = \lceil \sqrt{N_h} \rceil, \quad (12)$$

$$nRows = \lceil \frac{N_h}{nCols} \rceil, \quad (13)$$

where $\lceil . \rceil$ is the ceil function. Rows here correspond to the longitude dimension of the Earth-grid, whereas columns correspond to latitude. Despite the Earth-fixed grid is hexagonal, square blocks still can be applied by decomposing the hexagonal grid in two slightly shifted square grids. Let us numerically exemplify this hop index assignment process. For $N_h$=64, blocks of 8x8 are formed, meaning that each beam index is repeated every 9 beams in both latitude and longitude dimensions. For $N_h$=56, the blocks would be 8x7. For $56 < N_h < 64$, the decomposition in products of two integer numbers results in non-uniform block dimensions, yielding to reduced distances between active beams in one of the two dimensions. For instance, $N_h = 61$ only admits a decomposition in 61x1 blocks. Therefore, for these cases we stick to the 8x8 blocks though it is not possible to repeat hop indexes according to a rectangular pattern. In this case, we fill the first block starting from $ih = 0$ and reusing indexes until the block is fully filled. Then, the next block starts with the number that follows the last index of the previous block. We repeat this operation until the last index of a block is $ih = N_h - 1$. For instance, for $N_h = 61$, we would need 8 blocks of 8x8. The resulting 64x8 block is then regularly repeated along the Earth grid. This procedure is exemplified in Figs. 3-5 for $N_h = 61$.

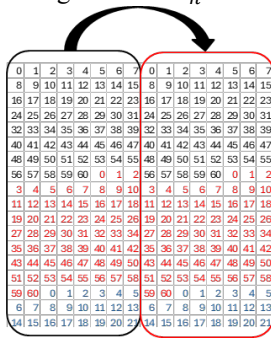

**FIGURE 3.** Example of hop index assignment for $N_h$=61 and beam layout design B. First 8 blocks of 8x8 are filled and concatenated vertically (black rectangle only showing two entire blocks), then the 8 blocks are repeated through the whole coverage (red rectangle, only one repetition on the horizontal dimension is shown).

Fig. 3 shows a portion of the hop index assignment process; Fig. 4 illustrates the beam footprints associated to $ih = 0$, which are simultaneously illuminated within the same hop; and Fig. 5 depicts the quasi-square patterns corresponding to blocks of grid points assigned to consecutive hop indexes from $ih = 0$ to 60. Note that in each hop, only one beam per each of these blocks is illuminated.

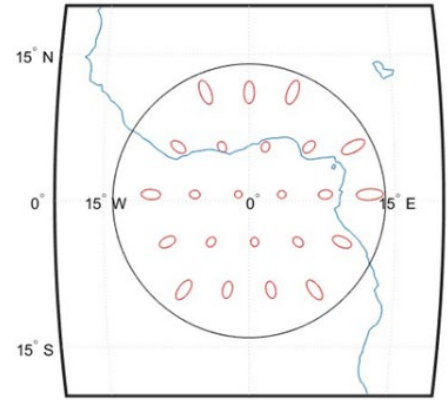

**FIGURE 4.** -3 dB beam contours of beams assigned to $ih = 0$ for $N_h$=61 and beam layout design B.

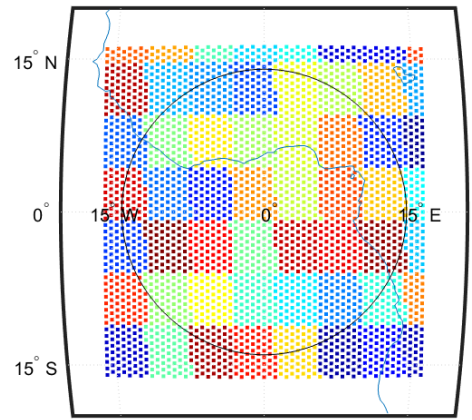

**FIGURE 5.** Hop index assignment for $N_h$=61 and beam layout design B. Grid points indicated by the same color are assigned to the same block of consecutive hop indexes from $ih = 0$ to $ih = 60$.

## VI. NUMERICAL EVALUATION

This section presents a numerical comparison of the beam layout designs discussed in Section IV, considering the hop index mapping described in Section V. First, the system level parameters are introduced. Next, the evaluation procedure and the numerical key performance indicators (KPIs) are described. Finally, the results are presented.

### A. SYSTEM PARAMETERS

Table II summarizes main simulation parameters. It considers a LEO satellite orbiting at 1300 km providing 5G services on a 200 MHz channel at Ka-band (i.e. FR2). The target coverage area corresponds to the Earth region in which the users see satellites with an elevation angle of at least 30º. It translates to a satellite tilt angle from Nadir of 46º. The downlink antenna consists of 512 radiating elements in an approximated circle shape with a Hexa-Deca structure and inter-element spacing of $0.65\lambda$. A per-radiating element power constraint is considered with an available power per element of 65 mW.

TABLE II
SIMULATION PARAMETERS

| Parameter | Value | Comments |
|---|---|---|
| Orbit height | 1300 km | |
| UE minimum elevation | 30º | |
| Minimum antenna beamwidth | 3.64º | At Nadir, averaged over φ |
| $\theta_{max}$ | 43.4º | Considering antenna beamwidth |
| $b_r(0)$ | 41.3km | |
| $b_r(\theta_{max})$ | 138.6km | |
| Radiating element gain | 4.7 dBi | |
| Frequency | 20 GHz | |
| Bandwidth | 200 MHz | |
| Available power | 65 mW | Per antenna element |
| Atmospheric loss | 0.5dB @elev=90°, 0.71dB @elev=45°, 1.1dB @elev=30° | Linear interpolation used for angles in between |
| Receiver G/T | 11dB @elev=90°, 10dB @elev=45°, 8dB @elev=30° | Linear interpolation used for angles in between |

### A. KPI AND EVALUATION PROCEDURE

The KPI under evaluation is the number of time resources required to illuminate 95% of the coverage area with a minimum SINR of $SINR_{th} = 3\,dB$. Time resources here correspond to the number of hops required, $(N_h)$, as discussed in Section V. Targeting 95% is a reasonable choice, though the design and evaluation method proposed can be used for 99% or more challenging targets. The minimum SINR value is selected according to the SNR requirements in Table I and considering a margin of at least 7 dB for rain or shadow fading.

The numerical evaluation is carried out adopting the following procedure:
1) Design the beam layout according to Section IV.
2) Set a dense uniform grid on Earth to sense the received SINR. A 32.000 points hexagonal grid with distances between points of 15 km is constituted.
3) Assign each point on the sensing grid to the closest beam footprint according to the geodetic distance between the sensing grid point and the beam footprint centers.
4) Calculate the SINR in the sense grid points associated to beams assigned to hop index $ih$, which are the only ones simultaneously illuminated.
5) Repeat Step 4 for all hop indexes.
6) If the percentile 5 of the SINR over the whole coverage area is below 3 dB, then increase $N_h$ and repeat Step 4 and 5 until $SINR= 3dB$ is achieved

This procedure is carried out for 9 different designs introduced in Section IV: Design A; Design B; two versions of Design C with $\alpha = 0.5, 0.75$ (C1 and C2, respectively); and 5 cases of Design D parametrized by $d_g = \sqrt{3}r_\theta$, where $r_\theta$ is the distance between the projection on Earth of a tilt angle $\theta$ and the subsatellite point. Designs D1-D5 correspond to $\theta$=4.8º, 6º, 7º, 7.5º, 9º, respectively. Note that design B corresponds to $\theta$=3.64º. The total number of beams for each design is included in Fig. 12, as part of the results.

### B. RESULTS

Let us start assessing the beam widening procedure of Section IV.A. Fig. 6 depicts the pattern cuts of the beam at Nadir for the different designs. For Design A, the required large beamwidth of 12.2º (note that a 3.64º beamwidth is obtained in case of no widening) is well approximated by the widening process despite a gain ripple, which was already expected. In contrast, Designs C1 and C2 require beamwidths much closer to the reference 3.64º. In these cases, the widening process is not so accurate and systematically provides beamwidths below the target ones. This result may be improved by further optimization though here we pragmatically stick to the results obtained by the closed-form expression. The main consequence is that in the central region of the coverage, the beam footprints do not cross at -3 dB as shown in Fig. 2 (Design C), but some spacing between footprints contours will appear. In any case, this spacing is always smaller than the observed using no-widening, as in Fig. 2 (Design D).

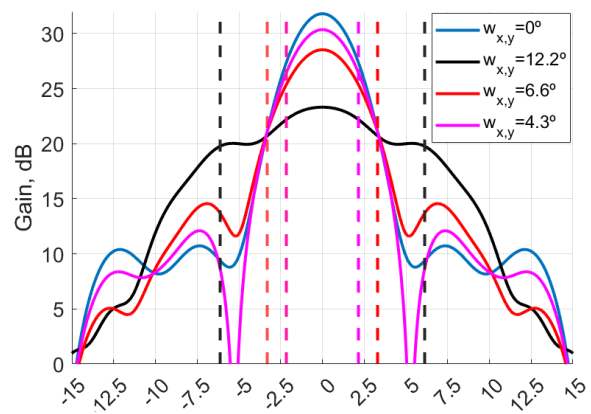

**FIGURE 6.** Pattern cuts at $\varphi = 0º$ for the beam at Nadir and for the different designs. Designs B and D1-5 have no widening so $w_x = w_y = 0º$, whereas $w_x = w_y$ =12.2º, 4.3º and 6.6º correspond to the widening required for Designs A, C1 and C2, respectively. Vertical dashed lines indicate the target beamwidth for Designs A (black), C1 (magenta) and C2 (red).

Fig. 7 further confirms the good approximation to the target uniform beams in the case of Design A. Note that beam footprints are plot corresponding to -5 dB beamwidth taking into consideration the observed gain ripple.

Let us now assess the beam power allocation solutions of Section IV.B. Fig. 8 and Fig. 9 plot the Cumulative Distribution Function (CDF) of the SINR received across the whole coverage obtained after step 5 in Section VI.B for Designs B and D2, respectively. For Design B, power allocation for SNR equalization allocates more power to beams closer to coverage edge, resulting in more balanced SINR levels that translate to clear improvements in the low percentile region. In contrast, in Design D2, this benefit disappears since lower SINR values are observed not only at

VOLUME XX, 2017

coverage edges but at the edge of beams close to Nadir, due to the increased distance between footprint grid points.

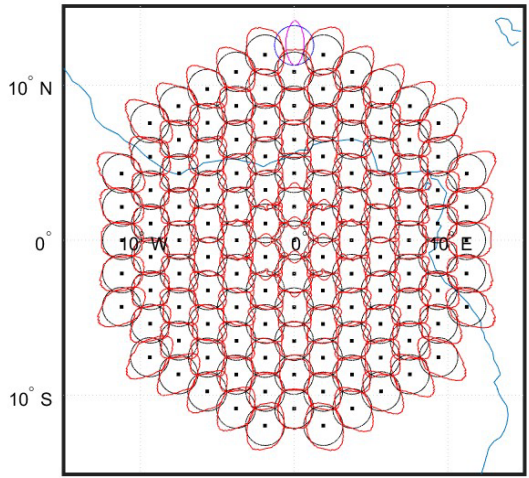

**FIGURE 7.** -5 dB beam footprints for Design A (red). Black circles depict the target uniform footprints. The magenta beam at the edge is used as a reference to calculate the target uniform pattern.

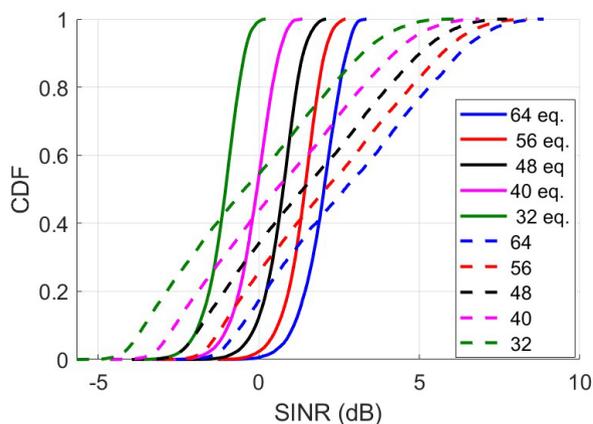

**FIGURE 8.** CDF of SINR over the whole coverage for Design B and $N_h$ increasing from 32 to 64. Equal power allocation and SNR equalization are compared.

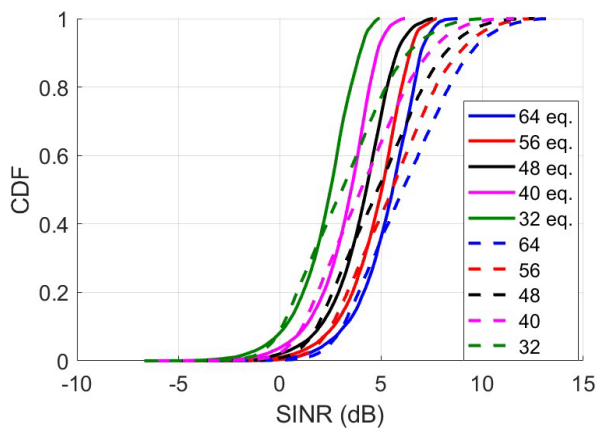

**FIGURE 9.** CDF of SINR over the whole coverage for Design D2 and $N_h$ increasing from 32 to 64. Equal power allocation and SNR equalization are compared.

The evaluation of Designs B and D2 is shown here as it exemplifies well the power allocation trade offs, but this procedure was repeated for all target designs concluding that power allocation for SNR equalization is beneficial for designs B, D1, C1 and C2. Designs D3-5 further increase the distance between grid points with respect to D2, so further degrade the benefits of the proposed power allocation scheme. The power allocation works for widened beam designs except for Design A, due to the large gain ripples around the main beam. Therefore, for the remainder of the evaluation, power allocation for SNR equalization will be applied to Designs B, D1, C1 and C2 while the rest will use equal power allocation. The results on the comparison for all designs are omitted for the sake of conciseness.

Fig. 8 and 9 depict also a general observed trend. As the value of $N_h$ is increased, the SINR increases since less simultaneous beams are transmitted in each hop. Fig. 10 and Fig. 11 depict the CDF of SINR for each design with the minimum $N_h$ value fulfilling the requirement of having 95% of the coverage with $SINR \geq 3\ dB$. Let us remind here that we focus on the minimum value, since it corresponds to the minimum number of hops required to sweep over the coverage area, so the minimum number of time resources devoted to broadcast common signaling. Although all curves cross at 3 dB and percentile 5, the SINR distributions are different. The equalized solutions (B, D1, C1, C2) provide more balanced SINR, whereas within the non-equalized, larger imbalances are observed as the number of total beams is decreased either through widening (A) or extending the beam cross over (D5).

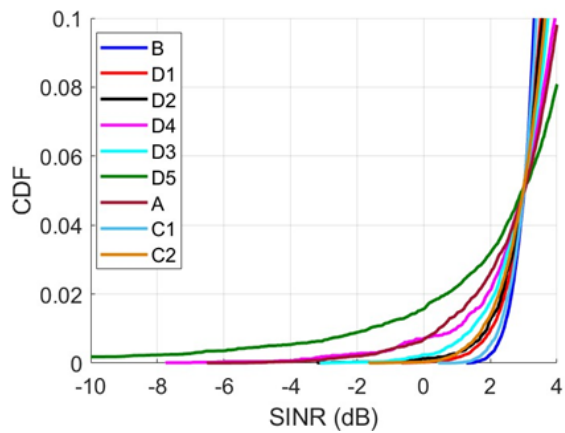

**FIGURE 10.** CDF of SINR over the whole coverage for the minimum $N_h$ value fulfilling the requirement of having 95% of the coverage with $SINR \geq 3\ dB$.

Fig. 12 shows the minimum $N_h$ value required to fulfill the target SINR. For designs based on maximum gain beams (B and D), an optimum configuration is found in D3, which minimizes the number of required hops. It uses a grid size $dg$ corresponding to a tilt angle of 7º from Nadir. Therefore, reducing the number of beams by enlarging the separation between them is beneficial up to point where the SINR at the edges of the beams close to Nadir is too low. In contrast, for beam widening solutions (A and C), the larger the reduction on the number of beams the better. Interestingly, the best designs of both techniques (D3 and A) provide similar results (62 for D3 vs 65 for A), but for a very different total number of beams. Fig.13 depicts the SINR distribution across the coverage area for both designs. In D3, the regions below 3 dB are the beam edges of beams at the coverage center and areas at the coverage edge. In contrast, in A the beam edges of beams at the coverage center present in general good SINR.

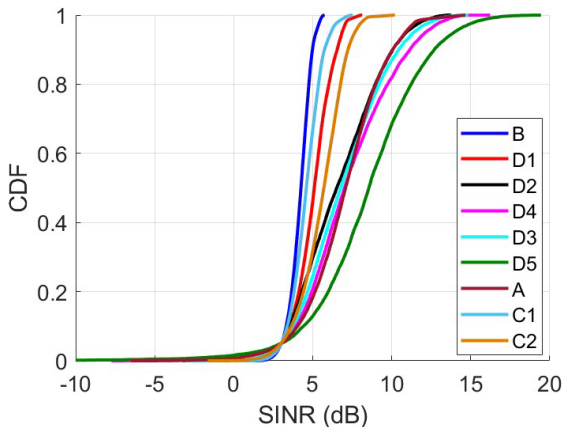

**FIGURE 11.** Zoom to low percentile region of Fig. 10

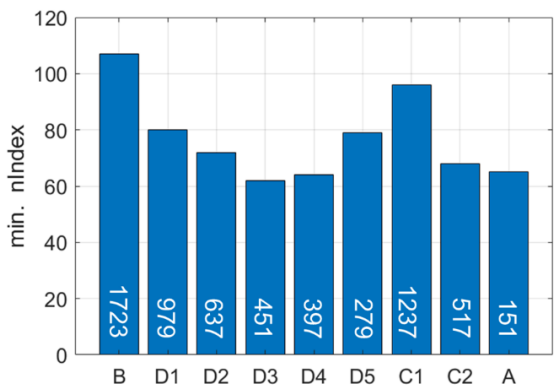

**FIGURE 12.** Minimum $N_h$ value fulfilling the requirement of having 95% of the coverage with $SINR \geq 3\ dB$. Values within the bars correspond to the total number of beam footprints within the coverage region.

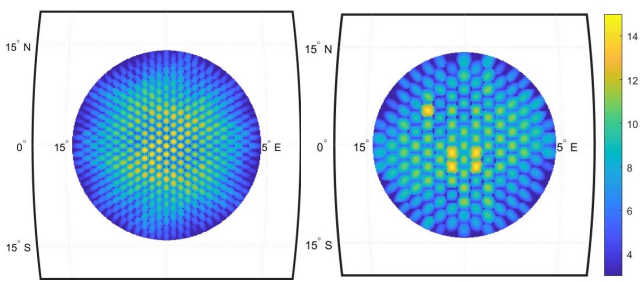

**FIGURE 13.** SINR distribution across the coverage area for the two most efficient designs: D3 (left) and A (right).

The differences in SINR levels for beams closely located is attributed to the different number of beams illuminated for each beam index. Note that, due to the regular beam index assignment using square blocks in Section V, and the circular shape of the coverage area, the number of beams assigned to the same beam index within the coverage area varies. For Design D3, the minimum and maximum number of beams simultaneously illuminated are 6 and 10, respectively, whereas for Design A, highest and lowest values are given by 1 and 4.

### VII. SCHEDULING OF COMMON SIGNALS

Section VI identified a beam hopping pattern minimizing the number of hops required to sweep over the whole coverage area while ensuring a minimum target SINR. The aim now is to map this beam hopping pattern to the 5G frame by defining the specific scheduling of the common signaling. In turn, it will define the hop duration and the number of full beam hopping sweeps to transmit all the common signaling across the coverage area.

The 3GPP standard already specified the timing of a beam sweep to transmit SSB burst across a cell in a half frame of 5 ms [7]. It supports up to 8 beams transmitting 8 SSB blocks in FR1, and up to 64 in FR2. For the remainder of the evaluation, we focus on FR2 assuming a subcarrier spacing of 120 kHz. In this case, the half frame contains 40 slots, 32 of them carrying two SSB beams each and the remaining 8 slots not carrying SSB signaling. The proposed approach is to co-schedule all common signaling within the slots devoted to SSB transmissions, so a single sweep may allow broadcasting all common signaling across the coverage. The target is to have one paging and one RA occasion per beam each SSB period, which by default can be either 20 ms or 160 ms. Note that in the case of 20 ms, SIB1 and SIB19 do not need to be allocated in each frame carrying SSB since their minimum periodicities are larger. Still, the available resources in half slot are very limited. To assess whether these resources are sufficient, we first derived the amount of resource elements required to transmit each common signal, as reported in Table III.

TABLE III
TIME/FREQUENCY RESOURCES REQUIRED

| Signal | Size in bits | PRBs | OFDM symbols |
|---|---|---|---|
| SSB | - | 20 | 4 |
| CORESET0/SS0 | - | 48 | 2 |
| CORESET1/SS1 | - | 132 | 1 |
| SIB1 | 1280 | 70 | 2 |
|  |  | 33 | 4 |
| SIB19 | 616 | 34 | 2 |
|  |  | 16 | 4 |
| Paging | 1280 (32 UE) | 70 | 2 |
|  |  | 33 | 4 |
|  |  | 22 | 6 |
| MSG2 | 490 (7 UE) | 27 | 2 |
|  | 560 (8 UE) | 15 | 4 |
|  | 630 (9 UE) | 11 | 6 |
| MSG4 | 1040 | 57 | 2 |
|  |  | 27 | 4 |
|  |  | 18 | 6 |
|  |  | 8 | 13 |

It assumes that the 3 dB SINR target allows using a Modulation and Coding Scheme (MCS) index of 6 [22],

which corresponds to QPSK modulation and code rate 449/1024 [23]. The required resources for SSB and CORESET0 are specified by 3GPP. We resort to SSB-CORESET0 multiplexing pattern 3 [7], so SSB uses 20 PRB and 4 symbols whereas CORESET0 is scheduled in the first two symbols of SBB and in the 48 PRB previous to the SSB. Therefore, a search space Type 0 (SS0) is already co-scheduled with each SSB block. Next subsections present three scheduling alternatives with increasing paging and RA capacities, followed by a comparative analysis.

### A. CO-SCHEDULING WITH SSB IN HALF SLOT
As anticipated, the first solution is to co-schedule all common signaling in the slots devoted to SSB transmissions, so the hop duration becomes half-slot. Therefore, the beams associated with up to 64 hop indexes $ih$ are sequentially illuminated within a half frame, since it supports up to 64 SSB blocks. According to the standard, the starting position of each SSB and SS0 within each slot varies with the SSB index (SSBI), being 4, 8, 2 and 6 for $SSBI \bmod 4 = 0,1,2,3$, respectively. As depicted in Fig. 14, when the SS0 starts at symbols 4 and 2 only two symbols after it remain available for common signaling accommodation. In view of this, we propose to set a CORESET1 occupying one OFDM symbol and the whole band used by a Search Space 1 (SS1) allocated at the first symbol of each slot. This enables PDSCH resources for RA and paging to be scheduled before the SS0. However, for SSB blocks starting at symbols 8 and 6 there is no room to include such SS1. The proposed solution is then to map the SSBI to different cell types. Cells type A will use only SSB blocks fulfilling $SSBI \bmod 4 = 0, 2$, and will map common search spaces Type 0 (SIB1) and 0A (SIB19) to SS0 and Type 1 (RA) and 2 (Paging) to SS1. In contrast, cells type B will use the rest of SSB blocks mapping all common search spaces to SS0. The drawback is that a cell with this configuration only supports up to 32 beams, and not the 64 provided by the standard.

Fig. 14 illustrates this scheduling concept showing the first two slots of a half frame transmitting 64 SSB blocks. It shows a reasonable allocation accommodating SIB1, Paging, MSG2 and MSG4 signaling in each half-slot corresponding to SSB. However, the number of MSG4, and the number of UE addressed in MSG2 and Paging varies with the SSBI. The worst case is observed for the half slot corresponding to SSB2, which only allows for 1 MSG4 and a paging message addressing 15 UE. In consecutive SSB periods, the resources used for SIB1, may be used for SIB 19 or for an extra MSG4 exploiting the larger required SIB1 periodicity. Assuming SSB periodicity of 20 ms and SIB1 and SIB 19 periodicities of 160 ms, in the worst case, the minimum MSG4 and paging capacity achieved in a beam would be 87.5 UE/s and 750 UE/s, respectively. Note that the allocation in Fig.14 is based on the default time resource allocation tables specified by 3GPP [23]. Customized tables would allow a better utilization of the resources within each slot, but still the room for MSG4 would be scarce.

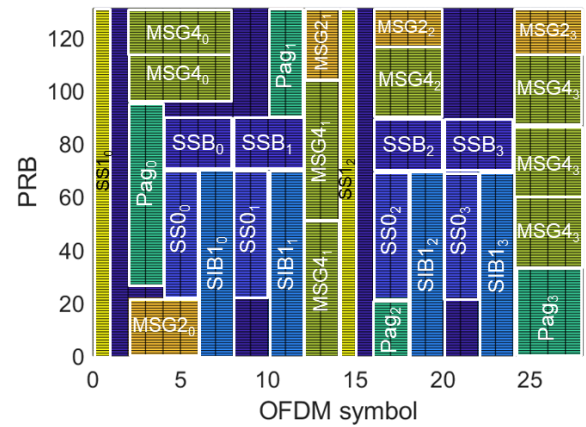

**FIGURE 14.** Allocation of common signaling assuming 64 SSB per frame so two SSB per slot.

### B. CO-SCHEDULING WITH SSB IN FULL SLOT
To increase paging and RA capacity and to reduce the differences between beams mapped to different SSBI, we propose to co-schedule all common signaling with SSB slots, but to transmit a single SSB per slot, thus enlarging the hop duration to one slot. In this case, the beams associated with up to 32 hop indexes $ih$ are sequentially illuminated within a half frame, since from the 64 SSB blocks supported by the standard, only 32 are used. As depicted in Fig. 15, this solution may allow the allocation of up to 9 MSG4, with paging messages addressing 32 UE, which is the maximum allowed by the standard. For SSB periodicity of 20 ms, the MSG4 and paging capacity are increased to 450UE/s and 1600 UE/s, respectively. Moreover, this scheme allows repetitions of SIB1 and SIB19 every 20 ms. Again, cells with this configuration only support 32 beams.

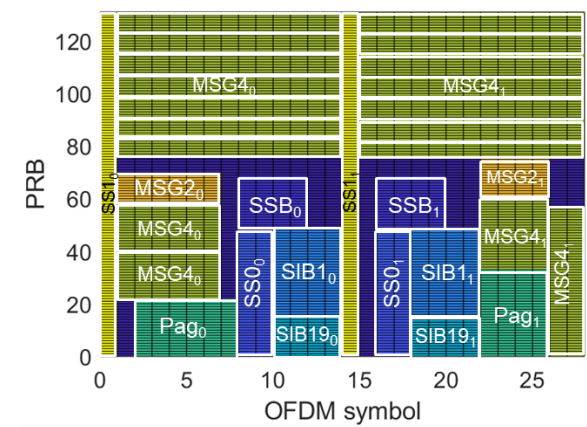

**FIGURE 15.** Common signaling allocation assuming a single SSB transmission per slot.

### C. EXTRA SWEEP AND SSB PERIODICITY OF 160 MS
In case it is needed, further capacity can be achieved by performing an extra sweep over the coverage not co-scheduled with SSB slots. The hop duration for this sweep can be set according to the required signaling capacity, though it makes sense to at least set it equal to one slot to have a clear increase of resources with respect to the previous solution. A reasonable option is to co-schedule SIB1, SIB19 and paging with SSB slots with a hop duration of half slot, leaving MSG2 and MSG4 for the extra sweep, with a SS1 indicating their allocations and a hop duration of 1 slot.

SSB periodicity of 160 ms constitutes a special case. The 3GPP specifies a maximum delay between MSG2 and the uplink grant for the transmission of MSG3 by the UE of 32 slots [24]. After its transmission, the UE expects to receive MSG 4 before 64 ms, which is the maximum time allowed for the Contention Resolution timer [24]. For LEO orbits, the periodicity of 160 ms exceeds the allowed delays between MSG2 and MSG4. Our recommendation would be to change the standard to support a single common signaling sweep in this case too. But within the current specification, the solution goes through an extra sweep for MSG4 performed within 64 ms after the transmission of MSG2. Hence, the proposal is to use the scheduling in Section VII.B but moving the MSG4 and its corresponding SS1 to a half frame not transmitting SSB bursts. The full slot solution of Section VII.B is chosen for the first sweep since it has enough space to schedule SIB1, SIB,19, paging and MSG2.

### D. COMPARISON

The comparison between the three scheduling solutions is made in terms of common signaling (CS) coverage ratio and efficiency. Coverage ratio was defined in 3GPP as the number of beam footprints broadcasting common signaling illuminated within a SSB period over the total number of beam footprints. The CS efficiency is defined here as the number of slots scheduling CS over the total number of slots within the SSB periodicity.

Fig. 16 shows a timing comparison between the three scheduling solutions assuming a SSB periodicity of 20 ms and two different beam hopping designs from Section VI: Design D3, which was the most efficient with $N_h = 62$; and Design B, with $N_h = 107$, which is treated as a benchmark since it does not involve any layout optimization. For $N_h = 62$, the three solutions can broadcast CS through the whole coverage within a SSB period, resulting in a coverage ratio of 100%. However, the half slot solution only requires a half frame to broadcast all CS in the 62 hops, whereas the full slot and the extra sweep schemes require 2 and 3 half frames, respectively. This translates to CS efficiencies of 80.6%, 61.3% and 41.9% for half slot, full slot and extra sweep, respectively. Without layout optimization, i.e. for $N_h = 107$, the number of half frames required to complete the beam hopping sweep over the whole coverage is larger. Consequently, only half slot and full slot solutions achieve a coverage ratio of 100%, but with CS efficiencies reduced to 66.3% and 33.1%, respectively.

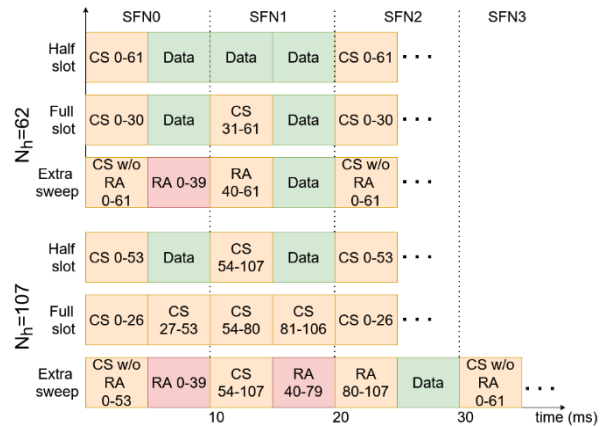

**FIGURE 16.** CS scheduling timing comparison for SSB periodicity of 20 ms and designs D3 and B in Section VI. Green subframes are devoted to data transmission, oranges are shared between CS and data transmission, and reds are exclusively used for CS.

The solution adopted by the 3GPP to increase both coverage ratio and CS efficiency has been to increase SSB periodicity to 160 ms though it may have a direct impact on the time required to acquire a cell or to transition from IDLE to active states. Fig. 17 repeats the timing analysis for this case but just focuses on the modified extra sweep proposed in Section VII.C. Now coverage ratios of 100% are easily satisfied even without layout optimization and CS efficiencies are largely increased. Specifically, they are 90.3% and 83.3% for $N_h = 62$ and $N_h = 107$, respectively, so an improvement of 7 percentage points is still observed for the layout optimization. Modifying the standard to allow a delay between MSG2 and MSG4 of more than 160ms, as recommended in Section VII.C, would further increase the efficiency for the optimized layout ( i.e. $N_h = 62$) up to 95.2%, since the beam hopping sweep devoted to MSG4 is not performed, but MSG4 are co-scheduled with the rest of common signals.

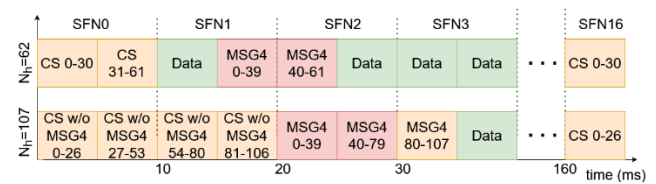

**FIGURE 17.** CS scheduling timing comparison for SSB periodicity of 160 ms and designs D3 and B in Section VI.

### E. BEAM TO CELL MAPPING

The 3GPP standard imposes two main restrictions that reduce the degrees of freedom on the mapping between satellite beams and NR cells. First, the maximum number of beams a cell can handle corresponds to the maximum number of SSB blocks within a frame, which as explained before can be up to 8 in Frequency Range 1 (FR1) and to 64 in FR2. Second, the SIB1 content distributed across the cell is the same for all beams, meaning for instance that common search spaces Type 0, 0A, 1 and 2 need to be the same across all beams. As described in Section VII.A, this has a direct impact if half slot scheduling is used.

Although a simple single beam to cell mapping would fit in this context, the proposal here is to use the maximum allowed number of beams per cell, so the number of cells is minimized. In this scenario, where each NR cell controls multiple beams, the legacy beam management can be used as baseline to handle user mobility across beams, avoiding more demanding handover procedures. Moreover, due to the beam overlapping, beam management can also be used to perform a beam switch in the event that a UE initially selects a wrong beam. Note that if the correct and wrong beams belonged to different cells, a handover would be necessary, which would increase signaling overhead. Let us also remark that a key distinction from terrestrial systems is that they can increase capacity through cell densification, deploying multiple base stations with independent time/frequency resource grids. In contrast, in EIRP-limited LEO systems, the time/frequency resource grids of different cells are not independent, since they need to be illuminated with the same limited set of beams. As a result, the capacity is limited by the number of active beams, regardless of the selected beam to cell mapping.

The proposed beam multiplexing methodology of Section V provides a straightforward beam to cell mapping, based on grouping blocks of beams assigned to consecutive hop indexes $ih$ to the same cell. Let us exemplify it for the beam layout design D3 with $N_h = 62$. Both half and full slot scheduling options support up to 32 SSB blocks per half frame, so cells group beams associated with consecutive hopping indexes either from $ih = 0$ to 31 or from $ih = 32$ to 61. The resulting cells show a quasi-rectangular pattern of 8x4 beam footprints, as depicted in Fig. 18. The extra sweep scheduling supports 64 SSB block per half frame so cells would group beams associated to consecutive hopping indexes from $ih = 0$ to 62, resulting in quasi-square patterns of 8x8 beams.

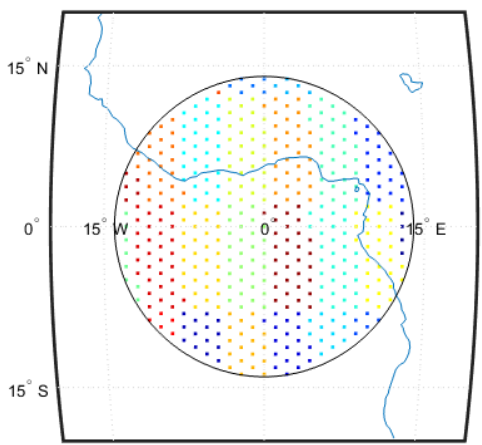

**FIGURE 18.** Beam to cell mapping for layout design D3 ( $N_h = 62$), for both half and full slot scheduling schemes. Each color indicates a different cell.

## VIII. CONCLUSIONS

We proposed an integral method for the design of beam management solutions targeting EIRP-limited LEO satellites, Earth-fixed beam operation and efficient 3GPP common signaling transmission. It included the derivation of the grid of beam footprints fixed on Earth, the design of the beamformers and beam power allocation and the planning of a regular beam hopping pattern considering the space, time and frequency resource allocation of the different 3GPP common signals. This method allowed the comparison of different grid and beam layout designs and the optimization of the design in terms of minimum resources required to transmit the common signaling over the whole coverage with a user received SINR over a given threshold.

A numerical evaluation using realistic system parameters allowed deriving remarkable conclusions. First, there is an optimum beam cross over when using phased array beamforming, which reduces the total number of beam footprints within the coverage while ensuring a minimum target SINR. Second, such optimum is not found for widened beams, for which the best solution is to widen them according to the beam footprint radius at coverage edge, producing quasi-uniform beams. Interestingly, phased array and widened beams provided very close results, but for a different number of total beams needed to illuminate the coverage area. Third, the scheduling of the common signaling over the optimized beam layouts assuming a by default SSB periodicity of 20 ms presents a tradeoff between the CS efficiency and the paging and RA capacity. For SSB periodicities of 160 ms, CS efficiencies are de facto increased but the beam layout optimization still provides non-negligible improvement.

The proposed method and the presented results have a direct application to the design of next 6G NTN LEO systems, especially if user positions are not known in advance so there is a need to illuminate the whole coverage. Combining this beam management for common signaling transmission with user centric beamforming solutions for data transmission will allow maximizing the capacity of these systems.

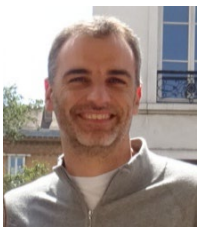

**XAVIER ARTIGA** (received the *Ingeniero Superior* (equivalent to M.Sc.) degree in telecommunications engineering from Universitat Politècnica de Catalunya (UPC), in July 2006. He joined CTTC, in September 2007, and was promoted to a Senior Researcher (R3A), in November 2020. He is currently the Technical Manager of TRANTOR (HE-SPACE). He is the co-author of more than 30 scientific publications. He participated in more than 20 research projects, including national, European, ESA, and industrially funded projects. His current research interests include reflectarray and metasurface reconfigurable antennas and multiantenna and Massive MIMO techniques for terrestrial and satellite systems. He received the Best Paper Award at ASMS/SPSC 2022 Conference.

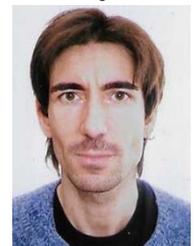

**MÀRIUS CAUS** (Senior Member, IEEE) received the M.Sc. and Ph.D. degrees in telecommunications engineering from the Universitat Politècnica de Catalunya, Barcelona, Spain, in July 2008 and December 2013, respectively. In 2018, he received the two-year Postdoctoral Juan de la Cierva Fellowship from the Spanish Government. He is currently a Researcher with the Centre Tecnològic de Telecomunicacions de Catalunya. He has participated in contracts with the industry and in several projects funded by European Commission, European Space Agency, and Spanish Ministry of Science. His main research interests include waveform design, signal processing for communications, and satellite communications.

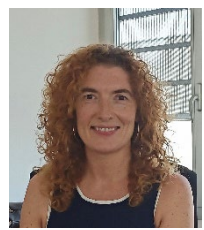

**ANA PÉREZ-NEIRA** is full professor at Universitat Politècnica de Catalunya in the Signal Theory and Communication department. Currently, she is the Director of Centre Tecnològic de Telecomunicacions de Catalunya, Spain. Her research is in signal processing for communications, focused on satellite communications. She has been the leader of over 35 projects and has participated in over 50. She is the author of over 80 journal papers and more than 400 conference papers. She is co-author of 7 books and 13 chapters, and 8 patents, all contributing to the advancement of signal processing and communications. She has been speaker at 70+ invited talks, guest editor for 10 special issues. She was Vicerector for Research at UPC (2010-13). She created UPC Doctoral School (2011). She is a recipient for the 2018 EURASIP Society Award. She is IEEE Fellow, EURASIP Fellow, member of the Real Academy of Science and Arts of Barcelona, and of the Real Academy of Engineering. She is awarded the ICREA Academia and has the Narcis Monturiol award by the Catalan government due to her research trajectory. She has been recognized in 2025 the Salvà Campillo Award for Outstanding Personality of the 30th Night of Telecommunications and IT.